\documentclass[runningheads]{llncs}
\usepackage[arrow, matrix, curve]{xy}
\usepackage[latin1]{inputenc}
\usepackage[ngerman, english]{babel}
\usepackage{tikz}
\usepackage{mathtools}
\usepackage{amssymb}
\usepackage{amsfonts}
\usepackage{amsxtra}
\usepackage{stmaryrd}
\usepackage{amssymb}
\usepackage{graphicx}
\usepackage{epstopdf}% To incorporate .eps illustrations using PDFLaTeX, etc.
\usepackage[bookmarksnumbered,colorlinks,bookmarks,citecolor=blue,linkcolor=blue,urlcolor=blue,breaklinks,linktocpage]{hyperref}% \usepackage[all]{hypcap}% This turns on the hyperlinks for \label{} and \ref{}, and others.
\usepackage{subfigure}% Support for small, `sub' figures and tables
\usepackage{xspace}
\usepackage{latexsym}

\newcommand{\I}{\mathrm{i}}
\newcommand{\E}{\mathrm{e}}
\newcommand{\D}{\,\mathrm{d}}

\newcommand{\pos}{\mathrm{pos}}

\newcommand{\Fig}[1]{Fig. \ref{#1}}
\newcommand{\Eq}[1]{Eq. \eqref{#1}}

\newcounter{commentcounter}

\begin{document}
\title{Quantum Tonality:  A Mathemusical Playground}
\titlerunning{Quantum Tonality:  A Mathemusical Playground}

\author{Peter beim Graben and Thomas Noll}
\authorrunning{Peter beim Graben and Thomas Noll}
\institute{
Bernstein Center for Computational Neuroscience Berlin, Germany \\
peter.beimgraben@b-tu.de \vspace{9 pt} \\
Escola Superior de M\'{u}sica de Catalunya \\
Departament de Teoria, Composici\'{o} i Direcci\'{o}, Spain \\ thomas.mamuth@gmail.com \\
}

\authorrunning{P. beim Graben and T. Noll}

\maketitle

\begin{abstract}
We adopt some basic ideas on quantum-theo\-retical modeling of tonal attraction and develop them further in an alternative direction. Fitting Gaussian Mixture Models (GMM) to the Krumhansl-Kessler (KK) probe tone profiles for static attraction opens the possibility to investigate the underlying wave function as the stationary ground state of an anharmonic quantum oscillator with a \emph{schematic Hamiltonian} involving a  perturbation potential. We numerically verify the fulfilment of the associated stationary Schr\"odinger equation and also inspect its excited states as a solution basis for the corresponding time-dependent Schr\"odinger equation. With their help we calculate the temporal evolution of any initial state. As an example, we study the dynamics of transpositions of the stationary KK wave function across Regener's line of fifths. This offers potential models for dynamic tonal attraction and also for the behavior of deflected key profiles within the Hamiltonian dynamics of a given tonality.

\end{abstract}

To appear in the \emph{Proceedings of the MCM Conference 2024}, Springer Verlag, Cham.

\section{Introduction}
In the present paper we experiment with a specific concept of \emph{tonality} that enters music theory from studies in cognition on the one hand and computer aided analy\-sis on the other. The psychological Krumhansl-Kessler (KK) probe tone profiles \cite{KrumhanslKessler82} for the major and minor keys describe \emph{static tonal attraction} of the various pitch classes in a tonal context and thereby reflect traces of enculturation within the Western musical mentality \cite{Huron06}, \cite{Temperley07}. Related profiles also result from statistical analyses of pitch class distributions in musical corpora and several authors have discussed connections between both kinds of data, e.g.~\cite{Arthur16}, \cite{Huron06}. In a more refined way, pitch class profiles are also used to represent the (fuzzy) pitch class content of local score segments and the investigation of their Discrete Fourier Transforms (DFT) provides valuable analytical insights through the tonal interpretation of the phases of prominent Fourier coefficients (see \cite{Amiot2016}, \cite{GrabenMannone20}, \cite{Yust2017a}). The KK profiles have also been interpreted as manifestations of pitch class hierarchies. For example, in his book \emph{Tonal Pitch Space}, Fred Lerdahl \cite{Lerdahl01} adopted the idea of embedded pitch hierarchies from Diana Deutsch und John Feroe \cite{DeutschFeroe81}, and analyzed tonal chord progressions in terms of such hierarchies. Moreover, melodic probe tone profiles and pitch class transition probabilities play an important role for \emph{dynamic tonal attraction} \cite{Huron06}, \cite{Woolhouse09}. These remarks indicate that pitch class profiles enter music theory in connection with various approaches to the investigation of tonality. Upon this wider background it is desirable to understand the implications of a more recent approach to the mathematical interpretation of the KK profiles, proposed by Reinhard Blutner and Peter beim Graben in \cite{Graben23}, \cite{GrabenBlutner19} using ideas from quantum theory.

Starting point of the present study is the observation in \cite[Fig.~1]{NollGraben22} that the deformed quantum cosine model $\psi: \mathbb{R}/ 2 \pi \mathbb{Z} \to \mathbb{C}$ over the continuous circle of fifths \cite{GrabenBlutner19} can be equivalently described by a Gaussian wave function $\psi: \mathbb{R} \to \mathbb{C}$ over Regener's line of fifths $\mathbb{R}$ as configuration space \cite{Regener73}, retaining the probe tone positions along the circle of fifths up to enharmonic equivalence, such that the KK profiles emerge as mixed quantum states, i.e. as weighted sums
\begin{equation}\label{eq:gmm3}
  p(x) = |\psi(x)|^2 = \alpha \cdot \rho(x - a) + \beta  \cdot  \rho(x - b) + \gamma  \cdot \rho(x - c)
\end{equation}
of three transposed instances of a Gaussian probability density function $\rho(x)$, where $a, b, c \in \mathbb{R}$ refer to the positions of the three components of the tonic triad. This finding leads to two important consequences. (\emph{i}) it links the quantum music approach to statistical machine learning and data science \cite{RussellNorvig10}, and (\emph{ii}) it allows the application of the quantum theory of harmonic and anharmonic oscillators \cite{Schrodinger26b},\cite{WeenyCoulson48}.

\section{Quantum Wave Functions for Gaussian Mixture Models}
This section summarizes the results from a longer companion article by the authors of the present submission \cite{GrabenNollIP}. First we recall that a Gaussian wave packet
\begin{equation}\label{eq:gausw}
  \psi(x) = \sqrt[4]{\frac{E_0}{\pi}} \E^{-\frac{E_0 x^2}{2}}
\end{equation}
with energy
\begin{equation}\label{eq:gmmen}
  E_0 = \frac{1}{2 \sigma^2}
\end{equation}
and variance $\sigma^2$ solves the stationary Schr\"odinger equation of the harmonic oscillator
\begin{equation}\label{eq:oss}
  - \psi''(x) + E_0^2 x^2 \psi(x) = E_0 \psi(x)
\end{equation}
for the ground state energy $E = E_0$ \cite{Schrodinger26b}.

Next, we consider a generalized Gaussian mixture model (GMM) [\Eq{eq:gmm3}] of the form
\begin{equation}\label{eq:gmm}
   p(x) = \sum_{k = 1}^N \alpha_k \, \rho(x - a_k)
\end{equation}
where
\begin{equation}\label{eq:normal}
      \rho(x) =  |\psi(x)|^2 = \sqrt{\frac{E_0}{\pi}} \E^{- E_0 x^2} = \frac{1}{\sqrt{2 \pi \sigma^2}} \E^{-\frac{1}{2} \left(\frac{x}{\sigma}\right)^2}
\end{equation}
is a normal distribution density with variance $\sigma^2 \in \mathbb{R}$, whose mean $0$ is shifted to a tone position $a_k \in \mathbb{R}$ in each corresponding summand for the given tonality, altogether constituting a tonal context $\{a_1, \dots, a_N \} \subset \mathbb{R}$ \cite{GrabenBlutner19}. The coefficients $\alpha_k \in \mathbb{R}$ are the respective weights in the convex linear combination \eqref{eq:gmm} with $\sum_{k = 1}^N \alpha_k = 1$. Such GMM play an important role in the domain of statistical machine learning \cite{RussellNorvig10}. Here, they serve as a mathematical model for the schematic traces of musical enculturation \cite{Huron06}, \cite{Temperley07}.

In the companion article \cite{GrabenNollIP}, we show that there exists a perturbation $W: \mathbb{R} \to \mathbb{R}$ of the harmonic oscillator potential such that the ground state solution $\eta_0(x)$ (for the same energy eigenvalue $D_0 = E_0$) of the anharmonic stationary Schr\"odinger equation \cite{WeenyCoulson48}
\begin{equation}\label{eq:mod}
  - \eta_n''(x) + \left [E_0^2 x^2 + W(x)\right ] \eta_n(x) = D_n \eta_n(x)
\end{equation}
has the probability density function $|\eta_0(x)|^2 = p(x)$ of the GMM \eqref{eq:gmm}. Without proof we mention here, that the perturbation potential is given by the formula
\begin{equation}\label{eq:pert}
   W(x)  = E_0^2 \frac{\displaystyle
     \sum_{k l} [(a_k - 3 a_l) x + a_l (2 a_l - a_k)] \alpha_k \alpha_l \, \rho(x - a_k) \rho(x - a_l) }
     {\displaystyle \sum_{k l} \alpha_k \alpha_l \, \rho(x - a_k) \, \rho(x - a_l)}
\end{equation}  where $\rho(x)$ and the parameters $E_0, \alpha_k$ and $a_k$ are the defining ingredients of the GMM (c.f. equations (\ref{eq:gmmen}, \ref{eq:gmm}) and \eqref{eq:normal}). The differential operator \begin{equation}\label{eq:schematic}
 H =  - \frac{\partial^2}{\partial x^2} + E_0^2 x^2 + W(x)
\end{equation}
is referred to as the \emph{schematic Hamiltonian} associated with the GMM \eqref{eq:gmm} reflecting a statistically acquired tonal schema \cite{Huron06}, \cite{Temperley07}.

\section{Fitting Gaussian Mixture Models for Static Tonal Attraction}
In the present paper we explore these results for the case of two concrete examples, namely a GMM \eqref{eq:gmm3} for the KK probe tone profiles in $C$ major and $C$ minor \cite{KrumhanslKessler82}. Table \ref{table:KKData} lists both data sets (major and minor) in a normalized form:
\begin{table}[htp]
\caption{Normalized KK probe tone profiles for C-major (top) and C-minor (bottom).}
\begin{center}
\begin{tabular}{|c|c|c|c|c|c|c|c|c|c|c|c|}
\hline
$D{\flat}$ & $A{\flat}$ & $E{\flat}$ & $B{\flat}$ & $F$ & $C$ & $G$ & $D$ & $A$ & $E$ & $B$ & $F{\sharp}$\\
$-\frac{5}{6} \pi$ & $-\frac{2}{3} \pi$ & $-\frac{1}{2} \pi$ & $-\frac{1}{3} \pi$ & $-\frac{1}{6} \pi$ & $0$ & $\frac{1}{6} \pi$ & $\frac{1}{3} \pi$ & $\frac{1}{2} \pi$ & $\frac{2}{3} \pi$ & $\frac{5}{6} \pi$ & $\pi$ \\ \hline
$0.$ & $0.0203$ & $0.0127$ & $0.0064$ & $0.2365$ & $0.524$ & $0.3764$ & $0.159$ & $0.1819$ & $0.2734$ & $0.0827$ & $0.0369$ \\ \hline
$0.0184$ & $0.1893$ & $0.3733$ & $0.1052$ & $0.1301$ & $0.4982$ & $0.2905$ & $0.1236$ & $0.1288$ & $0.0079$ & $0.0447$ & $0.$\\ \hline
\end{tabular}
\end{center}
\label{table:KKData}
\end{table}%

The note names $G{\flat}, D{\flat}, A{\flat}, E{\flat}, B{\flat}, F, C, G, D, A, E, B, F{\sharp}$ in the line-of-fifth order (with $G\flat$ added to the left side with the same values as for $F\sharp$) occupy the equidistant positions $x \in \{-\pi, -\frac{5}{6} \pi, \dots, -\frac{1}{6} \pi, 0, \frac{1}{6} \pi, \dots, \frac{5}{6} \pi, \pi\}$, centered around the note $C$ at the origin $x = 0$ from the original quantum model over the circle of fifths. We write $\pos(X) \in \mathbb{R}$ for the position of the note $X$ at the real line of fifths \cite{Regener73}. The original values are scaled such that the norms of the linear interpolations $p(x)$ yield unity. We fit several GMM onto the KK data in analogy with the approach in \cite{GrabenBlutner19}, i.e. we choose GMM of the form \eqref{eq:gmm3} with different constraints upon the parameters.

\begin{figure}[h! tbp]
\centering
\includegraphics[width= 10 cm]{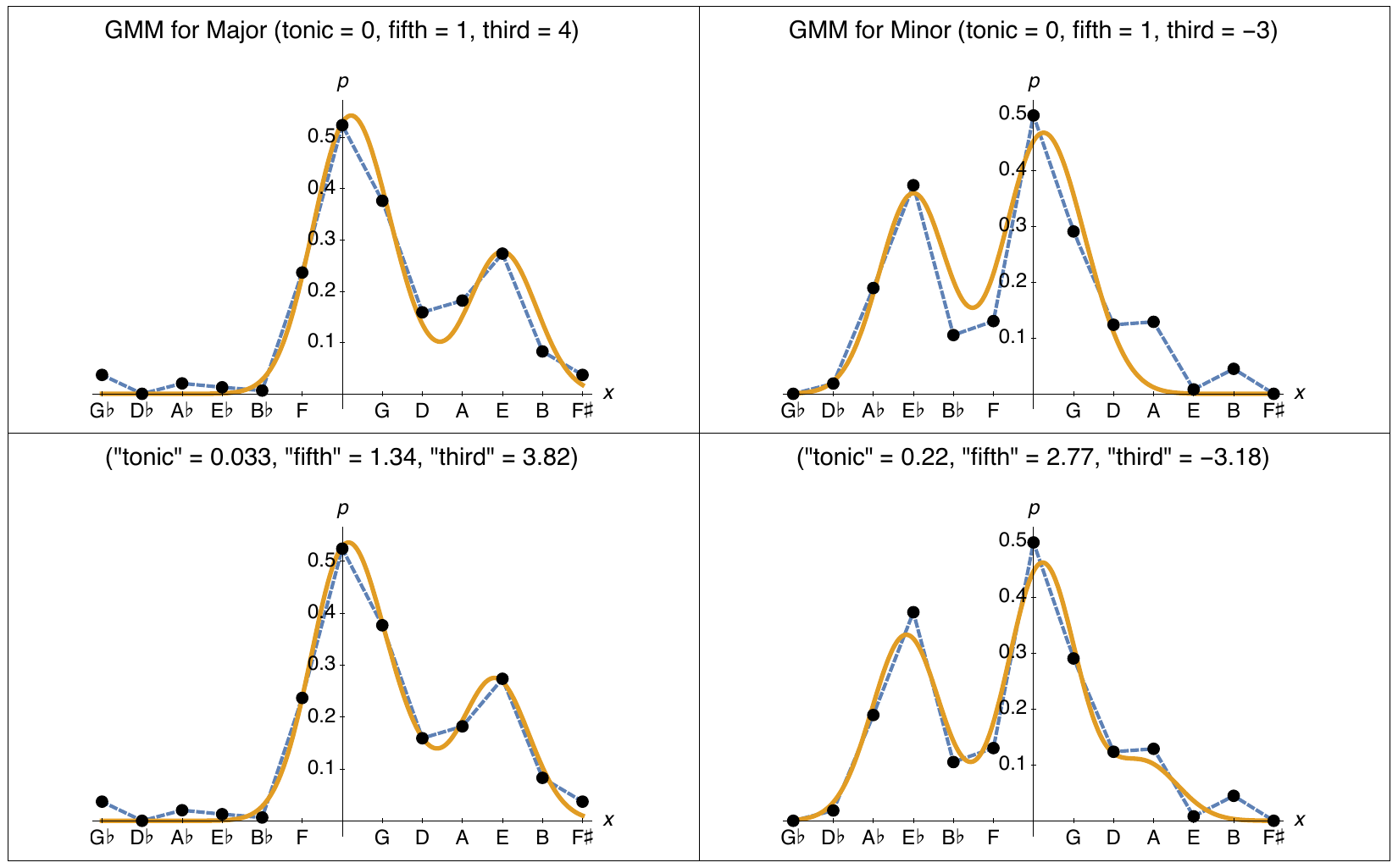}
\caption{Gaussian Mixture Models for the KK probe tone profiles for major (left) and minor (right). Top: means are chosen as triadic positions $a = 0 = \frac{\pi}{6}\pos(\text{tonic})$, $b = \frac{\pi}{6} = \frac{\pi}{6}\pos(\text{fifth})$, and  $c = \frac{\pi}{6}\pos(\text{third})$ ($= \frac{2 \pi}{3}$ or $-\frac{\pi}{2}$, respectively)  and the parameters $\alpha = 0.487468$, $\beta = 0.203587$ (major) and $\alpha = 0.405881$, $\beta = 0.194719$ (minor) are fitted, respectively. The correlation between the KK data and the corresponding values of the GMM are $0.989$ (for major) and $0.935$ (for minor). Bottom: means $a, b, c$ are fitted to the data as well.  The correlations are $0.995$ (for major) and $0.975$ (for minor).}
\label{fig:GMMPlots}
\end{figure}

The choice of the circle of fifths interval $[-\pi, \pi]$ motivates the inheritance of the ground state energy
\begin{equation}\label{eq:lerdahl}
    E_0 = \frac{36 \, \ln \, 2}{\pi^2}
\end{equation}
from the \emph{Lerdahl interpolation} in \cite{BlutnerGraben21}.\footnote{
    Note that we are able to confirm Lerdahl's speculation about a Hamiltonian principle of musical least effort \cite{Lerdahl01} in the present framework, as the ground state energy $E_0$ can be obtained from a corresponding variational principle.
}
In order to fit the most general GMM to the KK data we have five parameters: $a, b, c$ and $\alpha, \beta$ ($\gamma$ results from the convexity constraint $\alpha + \beta + \gamma = 1$). The locations of the three mean values $a, b, c$ are musically constrained through the assumption of the formative role of the tonic triads, i.e. $a = \pos(C), b = \pos(G), c = \pos(E)$ for the $C$ major context and $a = \pos(C), b = \pos(G), c = \pos(E_\flat)$ for the $C$ minor context. Beyond this it is also interesting to compare these choices with ``musically ignorant" fits to see how far they deviate from the musically motivated ones.\footnote{
    The authors are acutely aware of the danger of overfitting, in particular as the shape of $p$ evokes a connection between the elephant of Antoine de Saint-Exup\'{e}ry's \emph{The Little Prince} and \emph{John von Neumann's elephant}: ``With four parameters I can fit an elephant, and with five I can make him wiggle his trunk.'' \cite{jvn_elephant}
}

The results, displayed in \Fig{fig:GMMPlots}, indicate that in the case of the major key profile there is no significant difference between the two fits. In the case of the minor key profile, however, there are noteworthy differences in connection with the raised scale degrees $\sharp \hat{6}$ and $\sharp \hat{7}$ which deserve a separate investigation. For the scope of the present paper we focus on the case of the major key.\footnote{
    Fitting with a higher energy parameter  $E = 3.53406$ one can even achieve a correlation of $0.996338$, unfortunately, the match for the raised leading tone $\sharp \hat{7}$ is not yet convincing.
}

\section{Wave Functions of Static Tonal Attraction as Schematic Stationary Ground States}
We proceed with the $C$ major GMM \eqref{eq:gmm3} for tone positions $a = 0, b = \frac{\pi}{6}, c = \frac{2 \pi}{3}$ from the top left cell of \Fig{fig:GMMPlots}, with $\alpha = 0.487468$, $\beta = 0.203587$ and $\gamma = 0.3994$ defining our candidate for the associated schematic ground state wave function as $\eta_0(x) = \sqrt{p(x)}$.

Our next goal is to verify that $\eta_0(x)$ is indeed the stationary ground state of the schematic Hamiltonian (\ref{eq:schematic}) by solving the eigenvalue problem of the operator $H$. For this purpose, it is useful to take advantage of the Hermite basis of the harmonic oscillator, described by \eqref{eq:formal}.
\begin{equation}\label{eq:formal}
    \varphi_n(x) = \sqrt[4]{\frac{E_0}{\pi}} \frac{1}{\sqrt{2^n n!}} h_n(\sqrt{E_0} x) \E^{-\frac{E_0 x^2}{2}} \:,
\end{equation}
where $h_n(x)$ denotes the Hermite polynomial of order $n$. These Hermite states $\varphi_n(x)$ are solutions of the unperturbed Schr\"odinger equation \eqref{eq:oss} for the eigenvalues $E_n =  2 E_0 \left(n + \frac{1}{2}\right)$, i.e.
\begin{equation}\label{eq:hermite}
  - \varphi_n''(x) + E_0^2 x^2 \varphi_n(x)  = E_n \varphi_n(x) \:.
\end{equation}

When we apply the schematic Hamilton operator \eqref{eq:schematic} to a superposition state
\begin{equation}\label{eq:super}
    \psi(x) = \sum_n P_n \varphi_n(x)
\end{equation}
with coefficients $P_n = \left<\psi, \varphi_n \right> = \int_{-\infty}^{\infty} \psi(x) \varphi_n(x) \D x$, we obtain
\[
    H \psi(x) = \sum_n P_n \left[- \frac{\partial^2}{\partial x^2} + E_0^2 x^2 + W(x)\right] \varphi_n(x) = \sum_n P_n [E_n + W(x)]\varphi_n(x).
\]
The coefficients $P_n$ for eigenstates of the schematic Hamilton operator form eigenvectors of the infinite matrix $M = \left( e_{mn} + w_{mn}\right)_{m, n  \in  \mathbb{N}}$ where
\begin{equation}\label{eq:wmatx}
e_{mn} = E_n \delta_{mn} \quad \mbox{and} \quad
w_{mn} = \int_{-\infty}^{\infty} \varphi_m(x) W(x) \varphi_n(x)  \D x
\end{equation} (for details see the corresponding companion paper \cite{GrabenNollIP}). The potentials for the harmonic oscillator and its perturbation $W(x)$ in our concrete example (see \eqref{eq:pert}) are shown in \Fig{fig:Potentials}:

\begin{figure}[h! tbp]
\centering
\includegraphics[width= 10 cm]{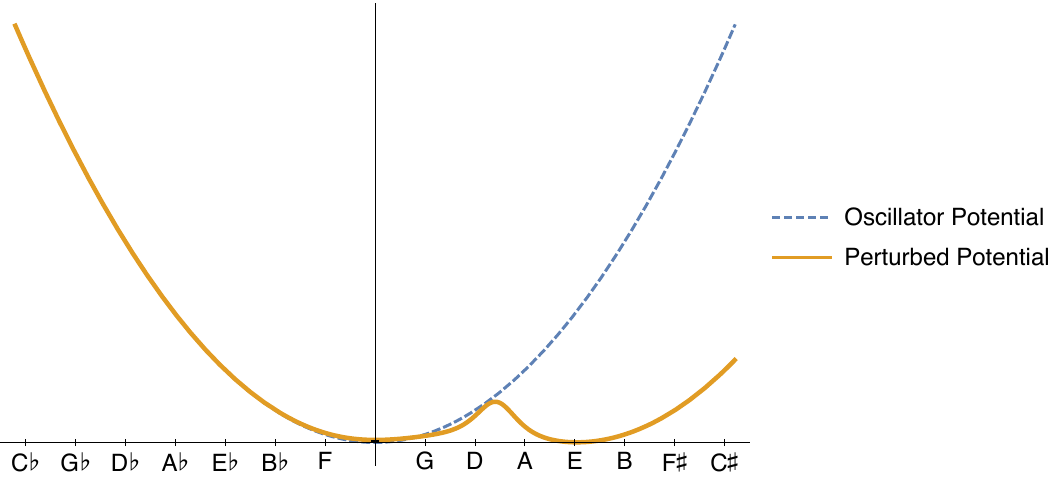}
\caption{The unperturbed potential of the harmonic oscillator with ground energy $E_0$ and perturbed anharmonic potential for the C major GMM $p(x)$.}
\label{fig:Potentials}
\end{figure}

For a numeric validation of the schematic Schr\"odinger equation, we calculate eigenvectors and eigenvalues of $M$ in growing (but finite) number of dimensions by truncating the infinite matrix after $N$ rows and columns: $M_N = \left( e_{mn} + w_{mn}\right)_{m, n  \le N}$.

\begin{figure}[h! tbp]
\centering
\includegraphics[width= 10 cm]{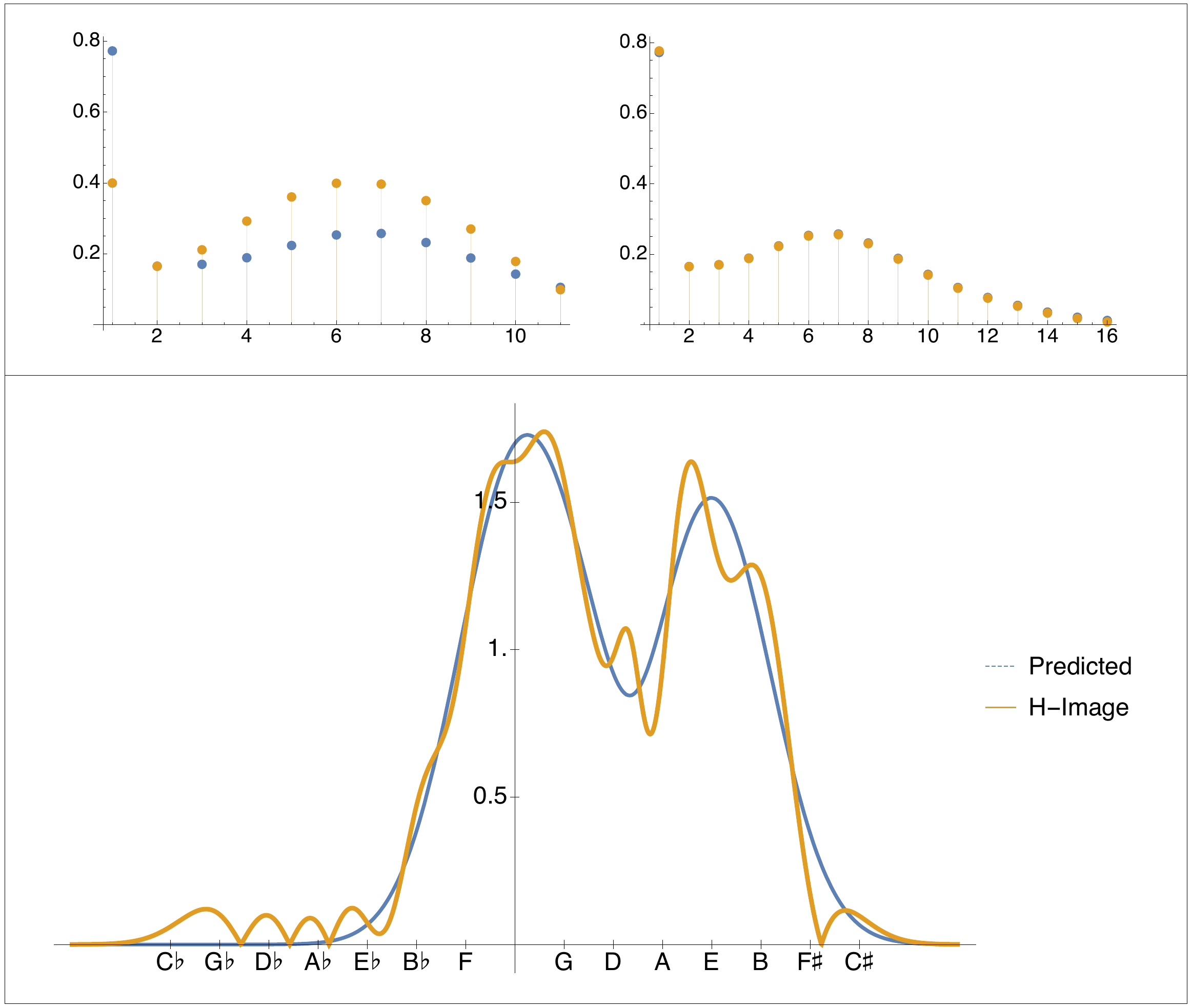}
\caption{Top: Comparison of the first $N$ coordinates $P_n$ from $\eta_0(x) = \sum_n Q_n \varphi_n(x)$ with the coefficients of the ``smallest" eigenvectors $v_{N} = (Q_0, \dots, Q_{N})$ of the truncated matrices $M_{N}$ for $N = 10$ (left) and $N = 15$ (right). Bottom: Comparison of the stretched wave function $E_0 \eta_0(x)$ with the action of the (truncated) Hamilton operator $H_{15}$ on the superposition $\sum_{n = 0}^{15} Q_n \varphi_n(x)$.}
\label{fig:CheckSchroedinger}
\end{figure}

The upper segment of \Fig{fig:CheckSchroedinger} shows list plots of the first $11$ (top left) and the first $16$ (top right) coordinates $P_n$ of $\eta_0(x)$ with respect to the Hermite basis \eqref{eq:hermite} and compares them with the coordinates of the ``smallest" eigenvectors of $M_{10}$ and $M_{15}$, respectively (i.e. the eigenvectors with the corresponding smallest eigenvalues). These eigenvalues ($2.617...$ and $2.531...$) are close to the Lerdahl ground state energy $E_0 = 2.528\dots$ \eqref{eq:lerdahl}, while one can still observe a clear deviation between the coefficients in the case of 11 dimensions (top left side of \Fig{fig:CheckSchroedinger}) the deviation already gets smaller in the case of 16 dimensions (top right side of \Fig{fig:CheckSchroedinger}). The plot in the lower segment of \Fig{fig:CheckSchroedinger} illustrates the fulfillment of the Schr\"odinger equation. The dashed curve shows the wave function $E_0 \eta_0(x)$, i.e. our predicted ground state stretched by the predicted energy eigenvalue $E_0$, while the continuous curve shows the image of the smallest eigenstate under the action of $H_N$, which is the approximation of the schematic Hamilton operator in terms of the truncated matrix $M_N$. For growing $N$ the little oscillations of the Hermite components become smaller and smaller and their superposition approaches the predicted wave function.

\section{Excited States of the Schematic Hamiltonian}
\begin{figure}[h! tbp]
\centering
\includegraphics[width= 8 cm]{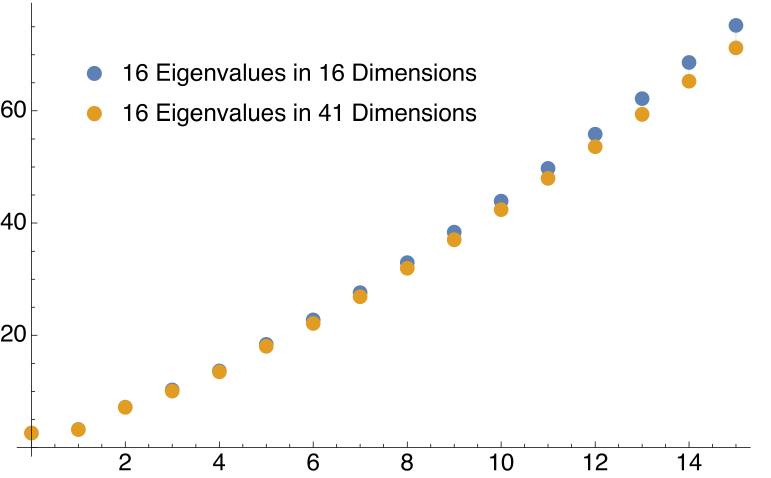}
\caption{The two list plots show the first 16 eigenvalues of $M_{15}$ and $M_{40}$, respectively}
\label{fig:EigenValues}
\end{figure}

In this section we explore an immediate consequence of the fact that we are in the possession of a Hamilton operator. Apart from the ground state we have an infinite sequence of higher energy eigenvalues and an infinite basis of corresponding excited stationary states, which form an orthonormal basis for the space of square-integrable complex functions over the line of fifths $\mathbb{R}$. In order to explore them, we may inspect their approximations from the sequence of eigenvectors of the truncated matrices $M_N$. Figure \ref{fig:EigenValues} suggests that --- for growing $N$ --- the eigenvalues of $M_N$ converge in quadratic mean the eigenvalues of $H$ from above, too.

\begin{figure}[h! tbp]
\centering
\includegraphics[width= 10 cm]{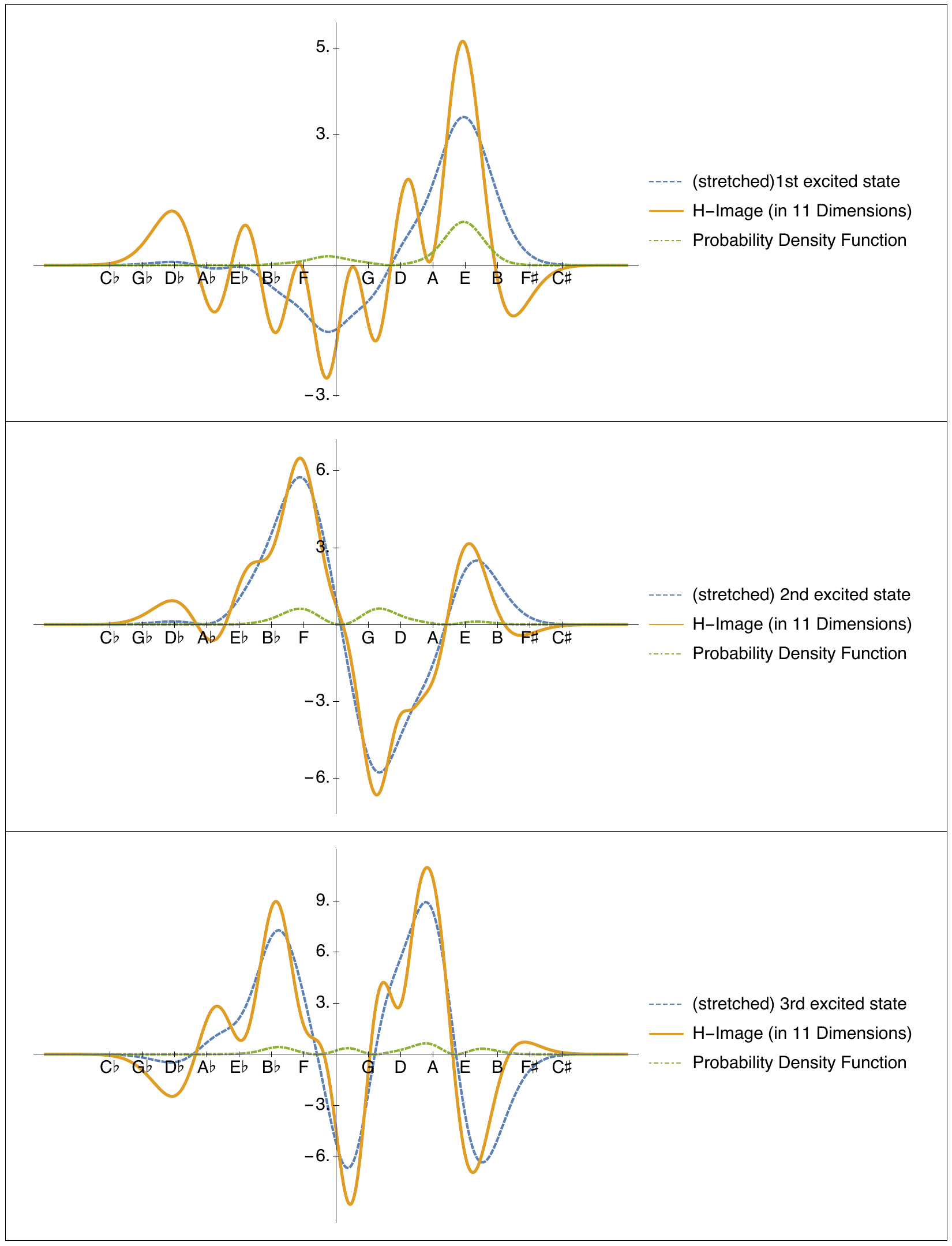}
\caption{Approximation of the 1st (top) , 2nd (middle) and 3rd (bottom) excited states of the schematic Hamilton operator $H$ by means of the 2nd, 3rd and 4th smallest eigenvalues and associated eigenvectors $v_{k} = (Q_{k,0}, \dots, Q_{k, 10})$ of the truncated matrices $M_{10}$. The dashed curves show the superpositions $E_k \sum_{n = 0}^N Q_{nk} \varphi_n(x)$ ($k = 1, 2, 3$), which are stretched by the eigenvalues $\tilde{D}_1 = 3.41579$, $\tilde{D}_2 = 7.29815$ and $\tilde{D}_3 = 11.1759$, respectively. The continuous curves show the actions of the (truncated) Hamilton operator $H_{10}$ on the superposition $\eta_k(x) = \sum_{n = 0}^{10} Q_{nk} \varphi_n(x)$. The dot-dashed curves show the associated probability density functions $|\eta_k(x)|^2$.}
\label{fig:ExcitedStates.pdf}
\end{figure}

Mathematically, we will now use the excited states in order to calculate the time-development of given initial states under the schematic Hamiltonian according to the time-dependent Schr\"odinger equation
\begin{equation}\label{eq:timesgl}
    H \Psi(x, t) = \I \frac{\partial}{\partial t} \Psi(x, t) \:.
\end{equation}

\section{Time Development of Deflected Key Profiles}
The excited states of the schematic Hamiltonian $H$, denoted $\eta_0, \eta_1, \eta_2, \dots$, form an orthonormal basis of the space of complex square-integrable functions on $\mathbb{R}$. The stationary ground state is the KK profile $\eta_0$ of static tonal attraction.

Suppose we have a representation of a wave function
\begin{equation}\label{eq:hcser}
      \psi(x) = \sum_k R_k \eta_k(x),
\end{equation}
whose coefficients are obtained by the orthogonal projections
\begin{equation}\label{eq:ortopro}
  R_k = \left<\eta_k, \psi\right> = \int_{-\infty}^{\infty}  \eta_k(x) \psi(x) \D x \:.
\end{equation}
Then, the solution of the time-dependent Schr\"odinger equation \eqref{eq:timesgl} is universally given by
\begin{equation}\label{eq:evolve}
    \Psi(x, t) = \sum_k R_k \E^{- \I D_k t} \eta_k(x)
\end{equation}
with the eigen energies $D_k$ of the schematic Hamiltonian \eqref{eq:schematic}.

Assuming that $\eta_0(x)$ represents the ground state of a given tonality (here $C$ major), it seems reasonable to associate other (major) tonalities with transpositions $\eta_0(x - a)$ of $\eta_0(x)$. However, regarding the underlying dynamics of the $C$ major tonality such transpositions are not autonomous representations of other tonalities at all. While $\eta_0(x)$ is stationary, $\eta_0(x - a)$ for $a \neq 0$ is not. The coherent states of the (unperturbed) quantum harmonic oscillator may serve as prime examples for our situation. They are the quantum-theoretical analogues for a deflected pendulum. We will therefore baptise the wave functions $\eta_0(x - a)$ (in the role as initial states) as \emph{deflected key profiles}. In order to explore their time developments we need to calculate the orthogonal projections
\begin{equation}\label{eq:ortopro}
  R_k = \int_{-\infty}^{\infty}  \eta_k(x) \eta_0(x-a) \D x \:.
\end{equation}

\begin{figure}[h! tbp]
\centering
\includegraphics[width= 10 cm]{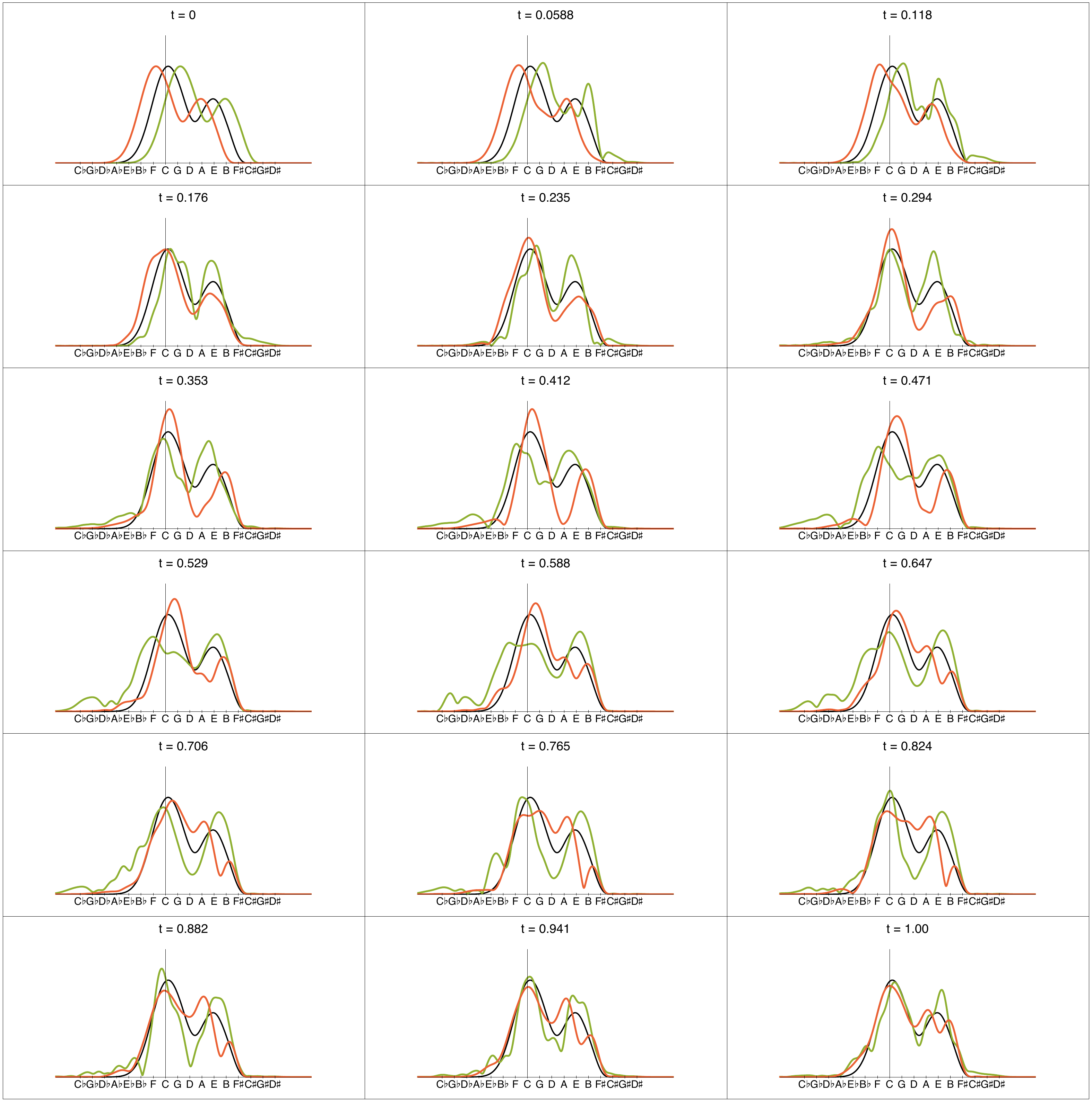}
\caption{Time Development of the deflected wave functions $|\eta_0(x - \frac{\pi}{6})|$ of one fifth-unit to the sharp side and functions $|\eta_0(x + \frac{\pi}{6})|$ of one fifth-unit to the flat side.}
\label{fig:SubdomDom}
\end{figure}

Figure \ref{fig:SubdomDom} gives a simultaneous flipbook-like trace (of the absolute values) of two time developments, namely of the fifth deflected wave functions $\eta_0(x - \frac{\pi}{6})$  in sharp-ward and $\eta_0(x + \frac{\pi}{6})$  in flat-ward direction. Both wave functions show the tendency to eventually arrive at the location of the stationary ground state. But in the details they behave quite differently and they never seem to match the precise shape again. From the numeric computation in 30 dimensions alone we cannot tell whether the deviations arise completely from the finite-dimensional approximation. A rough resemblance with the well-understood behavior of the coherent states of the harmonic oscillator cannot be dismissed out of hand.

\bibliographystyle{splncs04}

\addcontentsline{toc}{section}{References}

\end{document}